\documentclass[11pt]{article}
\usepackage{graphicx}

\newcommand{\BABARPubYear}    {04}

\newcommand{\BABARConfNumber} {25}
\newcommand{\SLACPubNumber} {10609}

\input babarsym

\def\Btag {\ensuremath{B_{\mbox{tag}}}\xspace}
\def\Bcp  {\ensuremath{B_{CP}}\xspace}

\def\KKKs      {\ensuremath{K^+ K^- \KS}\xspace}
\def\KKsKs     {\ensuremath{K^+ \KS \KS}\xspace}
\def\phiKs     {\ensuremath{\phi \KS}\xspace}
\def\sPlot{{\hbox{$_s$}{\cal P}lot}}

\long\def\inst#1{\par\nobreak\kern 4pt\nobreak
    {\it #1}\par\vskip 10pt plus 3pt minus 3pt}

\begin{document}
{\pagestyle{empty}

\begin{flushright}
\babar-CONF-\BABARPubYear/\BABARConfNumber \\
%\babar-PUB-\BABARPubYear/\BABARPubNumber \\
SLAC-PUB-\SLACPubNumber \\
%hep-ex/\LANLNumber \\
%BAD1007v7\\
%July 2004 \\
\end{flushright}

\par\vskip 5cm

% Title of the paper
\begin{center}
\Large \bf Measurement of {\boldmath \CP} Asymmetry in {\boldmath $ \Bz\to\KKKs$}\ Decays
\end{center}
\bigskip

\begin{center}
\large The \babar\ Collaboration\\
\mbox{ }\\
\today
\date{August 16, 2004}
\end{center}
\bigskip \bigskip

% Abstract
\begin{center}
\large \bf Abstract
\end{center}
We present preliminary measurements of the \CP\ asymmetry parameters and \CP\ content in
$\Bz \to \Kp\Km\KS$\ decays, with \phiKs\ events excluded.
In a sample of 227~M \BB\ pairs collected by the \babar\ detector at the \pep2\ 
\BF\ at SLAC, we find the \CP\ parameters to be 
	$S	=	-0.42 \pm 0.17 \pm 0.04$ and
	$C	=	 0.10 \pm 0.14 \pm 0.06$,
where the first error is statistical and the second is systematic.
Extracting the fraction of \CP -even final states from angular moments
$
	f_{even}~=~0.89 \pm 0.08 \pm 0.06,
$
and setting $C=0$, we determine
$
	\sin 2\beta ~=~ 0.55 \pm 0.22 \pm 0.04 \pm 0.11,
$
where the last error is due to uncertainty on the \CP\ content.

\vfill
\begin{center}

Submitted to the 32$^{\rm nd}$ International Conference on High-Energy Physics, ICHEP 04,\\
16 August---22 August 2004, Beijing, China

\end{center}

\vspace{1.0cm}
\begin{center}
{\em Stanford Linear Accelerator Center, Stanford University, 
Stanford, CA 94309} \\ \vspace{0.1cm}\hrule\vspace{0.1cm}
Work supported in part by Department of Energy contract DE-AC03-76SF00515.
\end{center}

\newpage
} % end of pagestyle{empty}

% Input author list file
\begin{center}
\small

The \babar\ Collaboration,
\bigskip

%% author list as of 02-Jul-2004 (609 authors)
%
B.~Aubert,
R.~Barate,
D.~Boutigny,
F.~Couderc,
J.-M.~Gaillard,
A.~Hicheur,
Y.~Karyotakis,
J.~P.~Lees,
V.~Tisserand,
A.~Zghiche
\inst{Laboratoire de Physique des Particules, F-74941 Annecy-le-Vieux, France }
% ...

\end{center}\newpage

% The body of the paper starts here
\section{INTRODUCTION}
\label{sec:Introduction}

In the Standard Model (SM) of particle physics, the decays $\Bz \to K^+K^-\KS$~\cite{charge} 
are dominated by $b\rightarrow s\bar{s}s$ gluonic penguin diagrams~\cite{sPenguin}. 
\CP violation in such decays arises from the Cabibbo--Kobayashi--Maskawa (CKM) quark-mixing mechanism~\cite{ckm}.

The time-dependent \CP\ asymmetry is obtained by measuring the proper time difference \deltat\ between
a fully reconstructed neutral $B$ meson (\Bcp) decaying into \KKKs, and the partially reconstructed 
recoil $B$ meson (\Btag).  The asymmetry in the decay rate ${\mbox{f}}_+({\mbox{f}}_-)$ 
when the tagging meson is a \Bz~(\Bzb) is given as
\begin{eqnarray}
{\mbox{f}}_\pm(\, \deltat)& = &{\frac{{\mbox{e}}^{{- \left| \deltat 
\right|}/\tau_{\Bz} }}{4\tau_{\Bz}}}  \, [
\ 1 \hbox to 0cm{}
\pm 
S \sin{( \deltamd  \deltat )}   
\mp 
\,C  \cos{( \deltamd  \deltat) }   ], 
\label{eq::timedist}
\end{eqnarray}
where $\tau_{\Bz}$ is the \Bz\ lifetime and \deltamd\ is the \Bz--\Bzb mixing frequency.
The parameters $C$ and $S$ describe the magnitude of \CP violation in the decay and
in the interference between decay and mixing, respectively.
In the SM, we expect $C=0$ because there is only one decay mechanism and direct \CP\ violation 
requires amplitudes with different \CP-violating phases.
If \KKKs\ decays proceed through a P(S)-wave leading to a \CP -odd~(even) final state, 
we expect $S=(-)\sin{2\beta}$, where $\sin2\beta=0.731 \pm 0.056$~\cite{Aubert:2002ic, Abe:2003yu}. 

However, contributions from physics beyond the SM could invalidate these predictions~\cite{Grossman:1996ke}.
Since $b\rightarrow s\bar{s}s$ decays involve one-loop transitions, they are especially sensitive to such contributions.
Recent results in decays of neutral $B$ mesons through the \phiKs\ intermediate state
are inconclusive due to large statistical errors~\cite{Abe:2003yt,Aubert:2004ii}.

A more accurate \CP\ measurement can be made using all decays to
\KKKs\ that do not contain a $\phi$ meson.
This sample is several times larger than the sample of \phiKs~\cite{Aubert:2003hz,Garmash:2003er,Aubert:2004ta}, but
the \CP\ content of the final state is not known.
In this paper we present measurements of the \CP\ content in \KKKs\ decays using an
angular-moment analysis and cross-check it with an isospin analysis~\cite{Garmash:2003er}.
We update our measurement of the \CP\ asymmetry parameters and extract the SM parameter $\sin2\beta$
with almost twice the statistics of the previous \babar\ result~\cite{Aubert:2004ta}.

\section{EVENT SELECTION}
\label{sec:event_selection}

This analysis is based on about 227 million \BB\ pairs collected
with the \babar\ detector~\cite{Aubert:2001tu} at the \pep2\
asymmetric-energy \epem\ storage rings at SLAC, operating on the $\Upsilon(4S)$
resonance.

We reconstruct $B$ mesons from $\KS\to\pi^+\pi^-$ and $K^\pm$ candidates.
Charged kaons are distinguished from pions and protons using energy-loss (\dedx) information in the tracking
system and from the Cherenkov angle and number of photons measured by 
the detector of internally reflected Cherenkov light (DIRC).
We accept $\KS\to\pip\pim$ candidates that have a two-pion invariant mass within $12$~\mevcc\ of the nominal \KS\ mass~\cite{pdg2004}, 
a decay length greater than three standard deviations, 
and an angle between the line connecting the $B$ and \KS\ decay vertices and the direction of \KS\ momentum vector less than 45~mrad.
The three daughters in the $B$ decay are fitted constraining their paths to a common vertex, and the $\pi^+\pi^-$ mass to
the nominal \KS\ mass.

In the characterization of the $B$ candidates we use two kinematic variables.
The energy difference $\Delta E = {E_B}^* - \sqrt{s}/2$ is reconstructed from the energy of the $B$ candidate ${E_B}^*$ 
and the total energy $\sqrt{s}$, both evaluated in the \epem\ center-of-mass (CM) frame.
The  \DeltaE\ resolution for signal \B decays is 18~\mev.
We also use the beam-energy-substituted mass $\mes = \sqrt{({s}/{2} + {\vec{p}}_{i} \cdot  {\vec{p}}_{B} )^{2}/{E^{2}_{i}} - {{\vec{p}}_{B}}^{\,2}}$,
where $( {\vec{p}}_{i}, E_{i} )$ is the four-momentum of the initial \epem\ system and 
${\vec{p}}_{B}$ is the momentum of the $B$ candidate, both measured in the laboratory frame.
The \mes\ resolution for signal \B decays is 2.6~\mevcc.\ 
We retain candidates with  $|\DeltaE|<200$~\mev and $5.2<\mes<5.3$~\gevcc.

The background is dominated by random combinations of tracks created in $\epem\to q\bar{q}~(q=u,d,s,c)$ continuum events. 
We suppress this background by utilizing the difference in the topology in the CM frame between jet-like $q\bar{q}$ events 
and spherical signal events.
The topology is described using the angle $\theta_T$ between the thrust axis of the $B$ candidate 
and the thrust axis of the charged and neutral particles in the rest of the event~(ROE)~\cite{Aubert:2001tu}.
Other quantities that characterize the event topology are two sums over the ROE:  $L_0=\sum|\vec{p_i}^*|$ and $L_2=\sum |\vec{p_i}^*| \cos{^2\theta_i}$,
where $\theta_i$ is the angle between the momentum $\vec{p_i}^*$ of particle $i$ and the thrust axis of the $B$ candidate.
Additional separation is achieved using the angle $\theta_B$ between the $B$-momentum direction and the beam axis.
After requiring $|\cos\theta_T| < 0.9$, these four event-shape variables are combined into a Fisher discriminant $\mathcal F$~\cite{Fisher:et}.

The remaining background originates from $B$ decays where a neutral or charged pion is missed during
reconstruction (peaking $B$ background).
We use samples of exclusive Monte Carlo (MC) events to model the signal and the peaking background,
and data sidebands to model the continuum background.

We suppress background from $B$ decays that proceed through a $b \to
c$ transition leading to the \KKKs\ final state by applying invariant
mass cuts to remove $\Dz$, $\jpsi$, $\chi_{c0}$, and $\psi(2S)$
decaying into $\Kp\Km$,\ and $\Dp$ and $\Ds$ decays into
$\Kp\KS$. Finally, to suppress \B decays into final states with pions,
we apply particle identification criteria to limit the pion
misidentification rate to less than 2\%.

\section{MEASUREMENT OF {\boldmath \CP} CONTENT}
\label{sec:cp_content}

\begin{figure}[!htb]
\begin{center}
\begin{tabular}{ll}
\includegraphics[height=5cm]{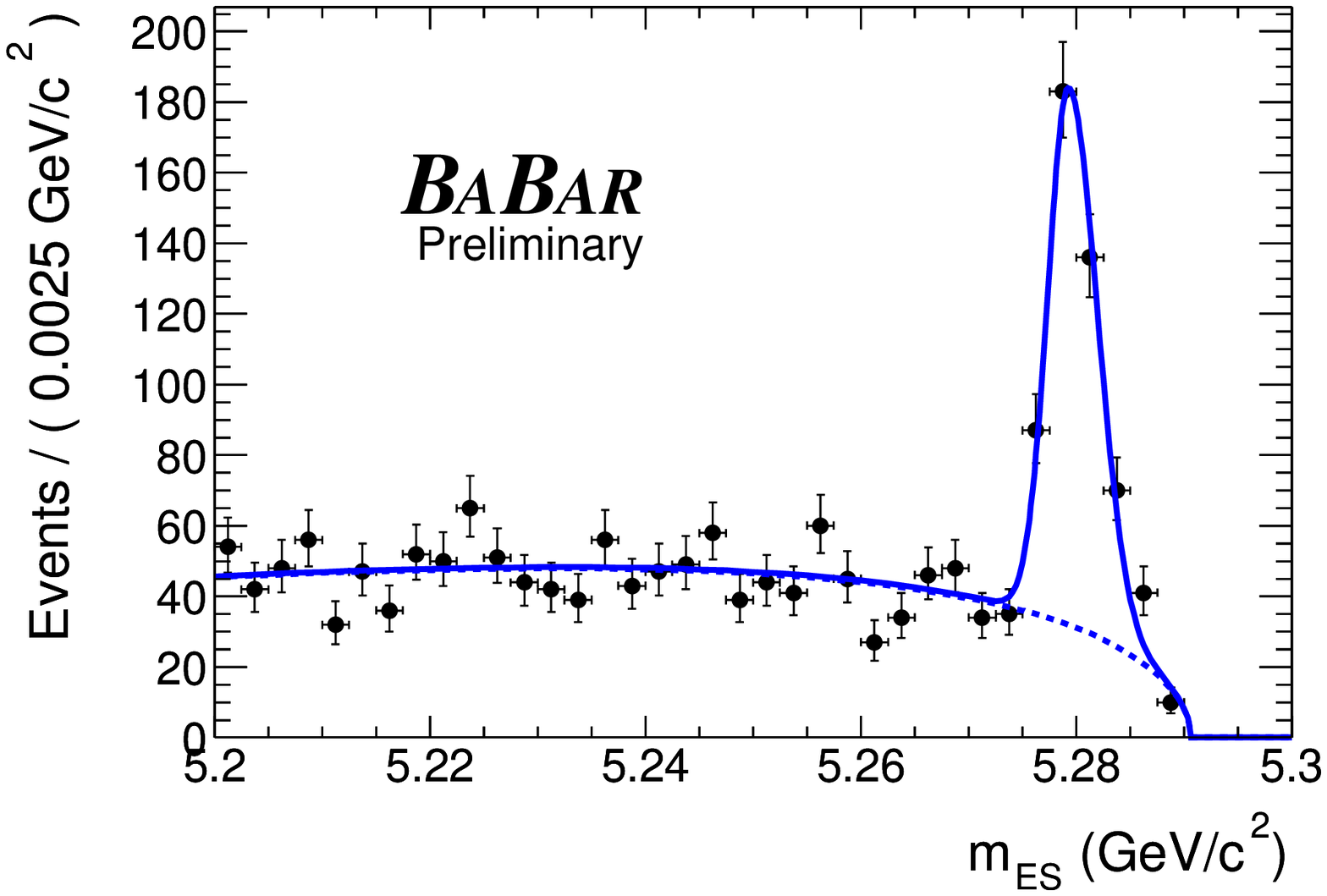} &
\includegraphics[height=5cm]{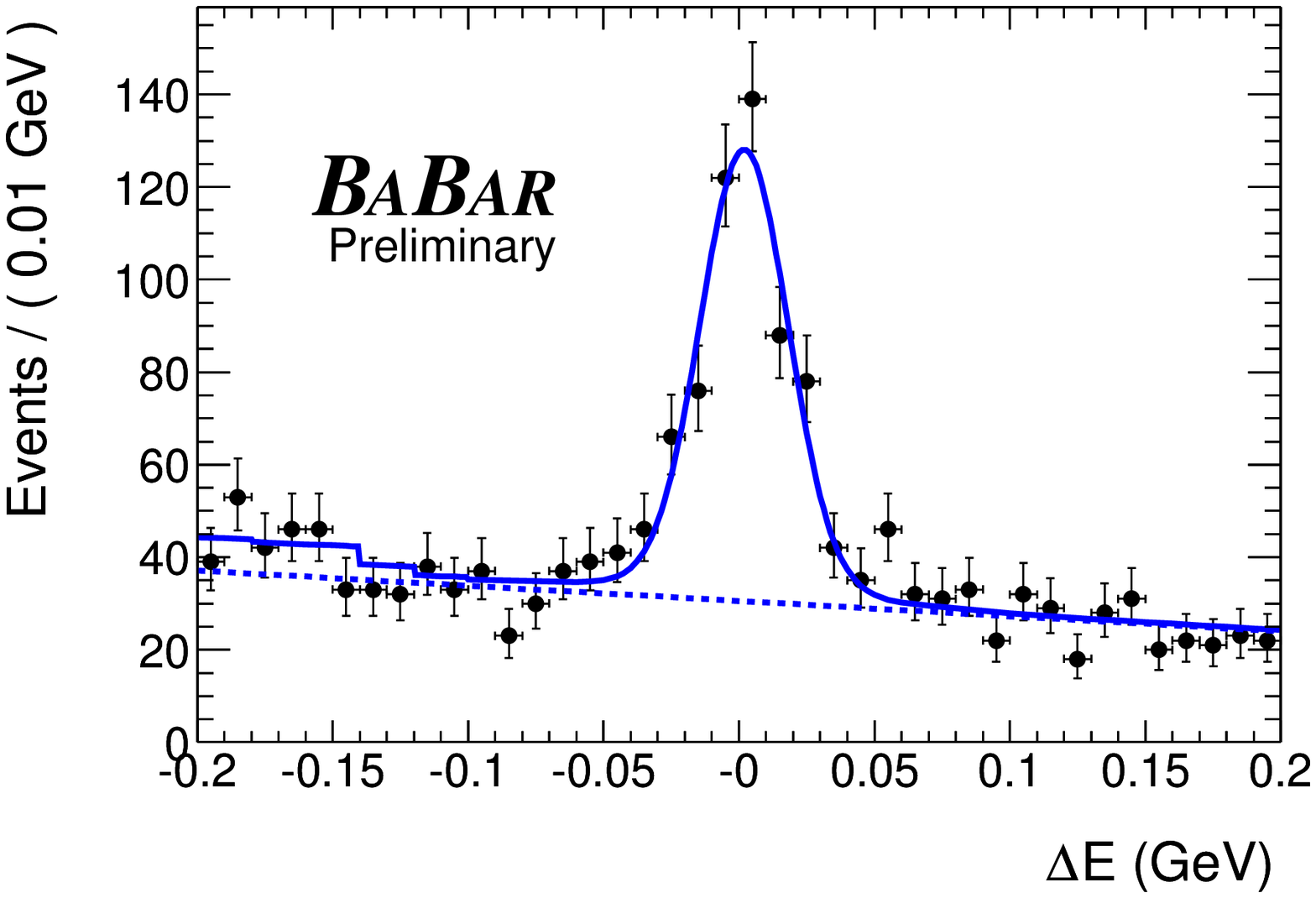} 
\end{tabular}
\caption{Projection plots of the \mes and  \DeltaE\ variables.
The points are data and the curves are projections from the likelihood fit for all events (full line) 
and continuum background (dashed line).
The signal-to-background ratio is enhanced with a cut on the signal probability.}
\label{fg::projection_plots}
\end{center}
\end{figure}

The extraction of the \CP\ content of the \KKKs\ final state is based on an unbinned extended maximum likelihood fit.
The likelihood function ${\mathcal L}$ is defined as: 
\begin{equation}
{\mathcal L} = \exp{\left(-\sum_{i}N_{i}\right)}
\prod_{j=1}\left[\sum_{i}N_{i}{\mathcal P}_{i,j}\right]
\label{eq::ml}
\end{equation}
where $j$ runs over all 27368~\KKKs\ events in the sample.
The probability density function (PDF) ${\mathcal P}$ is formed as  
${\mathcal P}(\mes) \cdot {\mathcal P}(\DeltaE) \cdot {\mathcal P}({\cal F})$.
Event yields $N_i$ for signal, continuum, and peaking $B$ background are floated in the fit.
We find a total of $525\pm30$ signal events in the entire \KKKs\ sample (including \phiKs), 
and show projection plots of the fit variables in Figure~\ref{fg::projection_plots}.

The \CP\ content is extracted using an angular moment analysis~\cite{Costa:1980ji}
which examines the distribution of the cosine of the helicity angle $\theta_H$
between the $K^+$ and \Bz\ directions in the $\Kp\Km$ center of mass frame. 
In this approach, we assume the decay rate  for a given $m(\Kp\Km)$ invariant mass  can be represented in terms of 
moments $\left < P_{l} \right  >$  of Legendre polynomials $P_l(\cos\theta_H)$ (see Appendix~\ref{sc::appendix})
\begin{eqnarray}
	|{\cal A}|^2 	&=& \sum\limits_{l} \left < P_l \right > \cdot P_l(\cos\theta_H),
\label{eq::Pl}
\end{eqnarray}
where $\cal A$ is the decay amplitude.
Since the dynamics of the decay is not known, we extract the moments by summing over all events 
\begin{equation}
\left < P_l \right > ~\approx~  \sum_j P_l(\cos\theta_{H,j}) ~{\cal W}_j / \varepsilon_j,
\label{eq::mom_sPlot}
\end{equation}
where $\cal W$ is the weight for event $j$ to belong to the signal decay.
We use the \sPlot\ background-substraction technique~\cite{Pivk:2004ty} to compute the
weights as ${\mathcal W}_j=\frac{ \sum_i V_{s,i} {\mathcal P}_{i,j} }{ \sum_i N_i {\mathcal P}_{i,j} }$,
where $V_{s,i}$ is the signal row of the covariance matrix obtained from the fit.
The efficiency $\varepsilon$ is evaluated from a high-statistics MC sample in $m(KK)-\cos\theta_H$ bins.
The covariance matrix for the moments is computed as 
\begin{equation}
\sigma_{ll'}^2  ~ \approx ~ \sum_j P_l(\cos\theta_{H,j})P_{l'}(\cos\theta_{H,j}) ~{\cal W}^2_j / \varepsilon^2_j,
\label{eq::sigmaPll}
\end{equation}
where the sum runs over events. 
Limiting ourselves to the two lowest partial waves, 
we can write the total decay amplitude in terms of 
the S-wave (\CP-even) and the P-wave (\CP-odd) amplitudes,
\begin{equation}
	{\cal A} ~=~ A_s P_0(\cos\theta_H) ~+~ e^{i\phi_p} A_p P_1(\cos\theta_H),
\label{eq::SP-waves}
\end{equation}
where $\phi_p$ is the relative phase between the partial-wave amplitudes $A_s$ and $A_p$.
It can be easily shown that the angular moments $\left < P_{0,2} \right  >$ give infomation on S- and P-wave strengths, and 
$\left < P_{1} \right  >$ arises from their interference.
Comparing Equations (\ref{eq::Pl}) and (\ref{eq::SP-waves}), 
we can relate the moments with the wave intensities and the total fraction of \CP -even events, $f_{even}$ as
\begin{eqnarray}
	A_s^2	 &=&  \sqrt{2} \left < P_0 \right > - \sqrt{\frac{5}{2}} {\left < P_2 \right >},  \\
	A_p^2	 &=&  \sqrt{ \frac{5}{2} } {\left < P_2 \right >}, \\
	f_{even} &=&  \frac{A_s^2}{A_s^2 + A_p^2} ~=~ 1 - \sqrt{\frac{5}{4}} \frac{ \left< P_2 \right >}{\left< P_0 \right >, } 
\label{eq::feven_moments}
\end{eqnarray}
where $A_s^2$ and $A_p^2$ are the S- and P-wave intensities, respectively. 
As a cross-check, we extract the fraction of \CP-even events by comparing 
the event rates of two isospin-equivalent channels~\cite{Garmash:2003er}:
\begin{eqnarray}
	f_{even}^{SU(2)} &=&	\frac{ \Gamma(\Bp \to \KKsKs)}{\Gamma(\Bz \to \KKKs)}
			=	\frac{ N_{\KKsKs} }{ N_{\KKKs} } 
				\frac{ \left < \varepsilon_{\KKKs} \right > }{ \left < \varepsilon_{\KKsKs} \right > }
				\frac{ \tau_{\Bz}}{\tau_{\Bp}}
\label{eq::feven_isospin}
\end{eqnarray}
where efficiencies $\left < \varepsilon \right >$ are averaged over the Dalitz plot and include branching fractions for $\KS\to\pip\pim$.

\section{MEASUREMENT OF {\boldmath \CP} ASYMMETRY}
\label{sec:cp_asymmetry}

The \CP\ asymmetry parameters are extracted from a \KKKs\ sample that excludes \phiKs\ decays by requiring $|m(K^+K^-) - m(\phi)| > 15 \mevcc$. 
We use the maximum-likelihood fit from Eq.~(\ref{eq::ml}), where the total PDF is formed as   
${\mathcal P}(\mes) \cdot {\mathcal P}(\DeltaE) \cdot {\mathcal P}({\cal F}) \cdot {\mathcal P}_c(\deltat;\sigma_{\deltat})$.
The time difference \deltat\  is extracted from the measurement of the separation \deltaz\ between the \Bcp\ and \Btag\ vertices, 
along the boost axis ($z$) of the \BB\ system.
The vertex position of the \Bcp\ meson is reconstructed primarily from the kaon tracks,
and its MC-estimated resolution ranges between 40--80\mum, depending on the opening angle and direction of the kaon pair.
The final resolution is dominated by the uncertainty on the \Btag\ vertex
which allows for a \deltat~(\deltaz) precision with an r.m.s.~of 1.1~ps~(180~\mum).
We retain events that have $|\deltat|<20$~ps and whose estimated uncertainty $\sigma_{\deltat}$ is less than 2.5~ps.
The \deltat\ resolution function is parameterized as a sum of two Gaussian distributions
whose widths are given by a scale factor times the event-by-event uncertainty $\sigma_{\deltat}$. 
A third Gaussian distribution, with a fixed large width, accounts for a small
fraction of outlying events~\cite{sin2b}.

Recoil side decay products are used to determine the flavor of the \Btag\ meson (flavor tag) and
to classify the event into seven mutually exclusive tagging categories~\cite{sin2b}. 
If the fraction of events in category $c$ is $\epsilon_c$
and the mistag probability is $\mistag_c$, the overall quality of the tagging, $\sum_c \epsilon_c (1-2w_c)^2$, is ($30.5 \pm 0.5$)\%.
Parameters describing the tagging performance and the \deltat\ resolution function for signal events are extracted from 
approximately 30,000 \Bz\ decays into $D^{(*)-}X^+\,(X^+ = \pip, \rho^+, a_1^+)$  flavor eigenstates. 

In the fit we float \CP\ asymmetry parameters, parameters describing the \deltat\ resolution function and tagging, event yields,
and the signal PDF parameters for \mes, \DeltaE, and the Fisher discriminant.

\section{SYSTEMATIC STUDIES}
\label{sec:Systematics}

In the measurement of the \CP-even fraction based on the angular moments we estimate a bias due to the
efficiency modeling from high-statistics MC events~(2.5\%). 
We do not find indication for the existence of higher moments $\left < P_l \right >$, $l=3\dots 6$ 
(see Figure~\ref{fg::moments} in Appendix~\ref{sc::appendix}), that could
arise from intermediate D-wave decays into \Kp\Km\ or decays proceeding through an $I=1$ resonance into $\Kpm\KS$.
Nevertheless, we estimate a systematic error from the D-wave by examining the $\left < P_2 \right >$ moment
in the $f_2(1270)$, $a_2^0(1320)$ and $f_2'(1525)$ mass region (1.1-1.7\gevcc) and assuming 
that $\left < P_2 \right >$ arises only from D-wave and S-D interference. 
Since the moment itself is consistent with zero, we assign a systematic error of 4\% based on the $\left < P_2 \right >$ error.
We also assign a systematic error due to potential contributions from decays proceeding through isovector resonances into $\Kp\KS$.
In addition to not being included in the angular moment analysis,
decays of charged $a_0^+(980)$, $a_2^+(1320)$ and  $a_0^+(1540)$ are not \CP\ eigenstates.
We set the error on the \CP\ content by counting events in $\Kpm\KS$ invariant mass regions under these resonances~\cite{pdg2004}.
We do not observe events in the $a_0(980)$ region, or a peak consistent with the $a_0(1540)$ resonance. 
We estimate events that could come from $a_2(1320)$ decays 
and conservatively include them into the systematic error (4.6\%).

The systematic errors on the time-dependent \CP -asymmetry parameters $(\sigma_S,\sigma_C)$  are estimated similarly to our previous
analysis~\cite{Aubert:2004ta}.
We account for the fit bias (0.02, 0.01), the presence of double CKM-suppressed decays in \Btag~\cite{Long:2003wq} (0.018, 0.053), 
the uncertainty in the beam spot and detector alignment (0.022, 0.012), and
the  asymmetry in the tagging efficiency for signal and background events (0.011, 0.014).
Other smaller effects come from the \deltat\ resolution and uncertainty 
on the \Bz\ lifetime and mixing frequency (0.006, 0.006). We use
$\tau_{\Bz}=1.536\pm0.014$~ps and $\deltamd =0.502\pm0.007~{\rm \ps}^{-1}$~\cite{pdg2004}.

%  sqrt(0.02*0.02+0.009*0.009),  sqrt(0.01*0.01+0.006*0.006)
%  sqrt(0.01*0.01 + 0.002*0.002 + 0.004*0.004),  sqrt(0.01*0.01 + 0.002*0.002 + 0.01*0.01)
%  sqrt(0.003*0.003 + 0.002*0.002 + 0.005*0.005)

\section{RESULTS}
\label{sec:Physics}

\begin{figure}[!tb]
\begin{center}
\includegraphics[height=10cm]{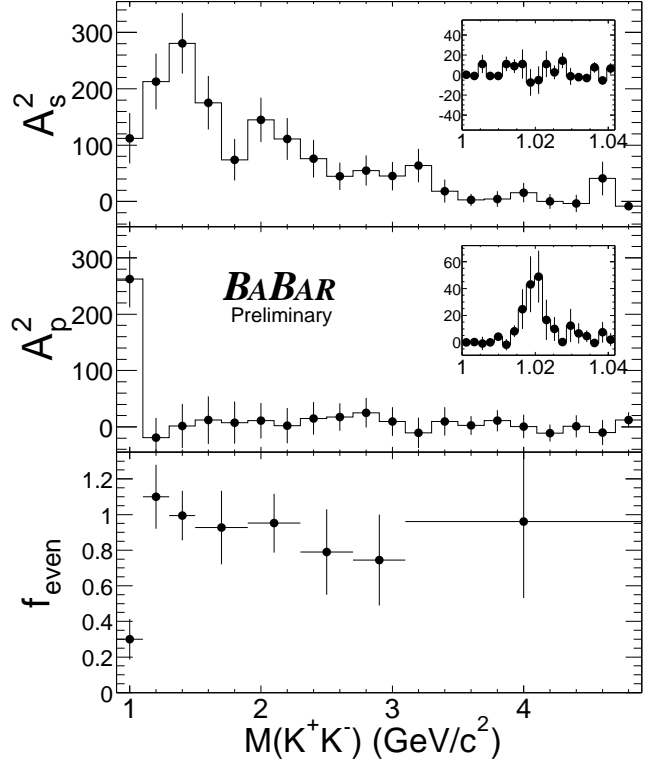}
\caption{Distributions of S- and P-wave intensities and \CP\-even fraction as a function of $\Kp\Km$ invariant mass. 
Insets show S- and P-wave intensities in the $\phi$ mass region. Events within 15~\mevcc of the nominal $\phi$ mass~\cite{pdg2004}
are removed from the \CP\ fit.}
\label{fg::waves}
\end{center}
\end{figure}

Distributions of the S- and P- wave intensities, and the \CP-even fraction as a function of $\Kp\Km$\ invariant mass are shown in Figure~\ref{fg::waves}. 
\phiKs\ events give a significant enhancement of P-wave decays in the first bin.
In the sample that excludes \phiKs\ events, we compute $\left < P_{0,2} \right >$ moments from the remaining events
and find the total fraction of \CP -even final states to be
$$
	f_{even} ~=~  0.89 \pm 0.08 \pm 0.06,
$$
where the first error is statistical  and the second is systematic.
We cross-check this result using the isospin approach of Eq~(\ref{eq::feven_isospin}).
We find  $452 \pm 28$ signal events in the \KKKs\ \CP\ sample with a total efficiency of $\left < \varepsilon \right > = (17.3 \pm 0.3)$\% 
and $208 \pm 18$ signal events in the \KKsKs\ sample with $\left < \varepsilon \right > = (9.8 \pm 0.8)$\%. 
This gives $f_{even}^{SU(2)}=0.75 \pm 0.11$ which is consistent with our nominal estimate.

% feven = 208./452.*0.173/0.098/1.086
% sigma = feven * sqrt( pow(28./452., 2)  +  pow(0.3/17.3, 2)  +  pow(18./208, 2)  +  pow( 0.8/9.8, 2) )

\begin{figure}[!htb]
\begin{center}
\includegraphics[height=10cm]{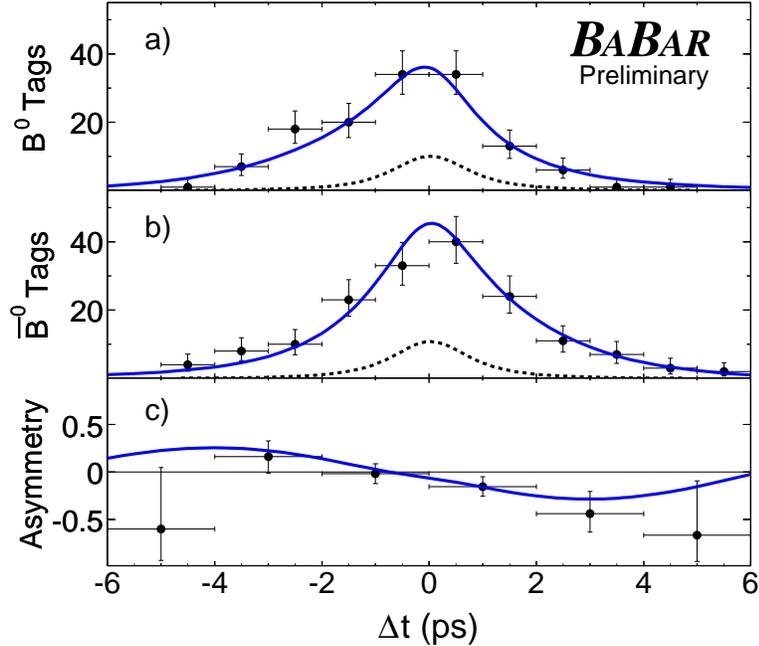}
\vspace{-8mm}
\caption{Distributions of $\Delta t$ for $\Bz \to \KKKs$ candidates with (a) \Bz tags and (b) \Bzb tags.
The solid lines refer to the fit for all events and the dashed lines correspond to the background contribution. 
The distribution of the raw asymmetry is shown in (c), where the solid line is obtained from the fit.
The signal-to-background ratio is enhanced with a cut on the signal probability.
\label{fg::dt}}
\end{center}
\end{figure}

The coefficients of the time-dependent \CP asymmetry
in $\Bz \to \KKKs$ decays (excluding \phiKs final states) are
determined to be
\begin{eqnarray}
	S	&=&	-0.42	\pm	0.17	\pm	0.04,	\nonumber \\
	C	&=&	 0.10	\pm	0.14	\pm	0.06.   \nonumber 
\end{eqnarray}
The \deltat\ distributions of events with \Bz\ and \Bzb\ tags, with projections from the likelihood fit superimposed, are shown in Figure~\ref{fg::dt}.
The fit procedure is verified with the \KKsKs\ sample, where we measure zero asymmetry,
and the $\jpsi\KS$\ sample where the results are consistent with previous measurements~\cite{Aubert:2002ic,Abe:2003yu}.

The presence of both P- and  S-wave decays in our \CP\ sample dilutes the measurement of the sine coefficient.
If we account for the measured \CP -odd fraction, we can extract the SM parameter $\sin 2\beta$.
Using the estimate of the \CP\ content based on the angular moments, and setting $C=0$ in the fit, we get  
$$
	\sin{2\beta} = -S/(2f_{even}-1) = 0.55 \pm 0.22 \pm 0.04 \pm 0.11,
$$
where the last error is due to uncertainty on the \CP\ content.

\section{SUMMARY}
\label{sec:Summary}

In a sample of 227 million \BB\ mesons, we have obtained preliminary measurements of the \CP\ content
and \CP\ parameters  in the \KKKs\ final state that excludes \phiKs\ decays.
From the distribution of the helicity angle in the $\Kp\Km$ frame, described in terms of moments of Legendre polynomials,
we extract the fraction of P-wave decays.
The result is consistent with our cross-check and previous measurements based on  isospin symmetry~\cite{Abe:2003yt,Aubert:2004ta}, 
and confirms the dominance of \CP-even final states.

We measure a time-dependent \CP\ asymmetry in $\Bz\ \to \KKKs$ decays at the $2.3\sigma$ level.
The obtained value for $\sin{2\beta}$  is consistent with the SM expectation and previous measurements in decays into the
\KKKs\ final state~\cite{Abe:2003yt,Aubert:2004ta}.

\section{ACKNOWLEDGMENTS}
\label{sec:Acknowledgments}

% Standard acknowledgments paragraph; must always be included.
We are grateful for the 
extraordinary contributions of our \pep2\ colleagues in
achieving the excellent luminosity and machine conditions
that have made this work possible.
The success of this project also relies critically on the 
expertise and dedication of the computing organizations that 
support \babar.
The collaborating institutions wish to thank 
SLAC for its support and the kind hospitality extended to them. 
This work is supported by the
US Department of Energy
and National Science Foundation, the
Natural Sciences and Engineering Research Council (Canada),
Institute of High Energy Physics (China), the
Commissariat \`a l'Energie Atomique and
Institut National de Physique Nucl\'eaire et de Physique des Particules
(France), the
Bundesministerium f\"ur Bildung und Forschung and
Deutsche Forschungsgemeinschaft
(Germany), the
Istituto Nazionale di Fisica Nucleare (Italy),
the Foundation for Fundamental Research on Matter (The Netherlands),
the Research Council of Norway, the
Ministry of Science and Technology of the Russian Federation, and the
Particle Physics and Astronomy Research Council (United Kingdom). 
Individuals have received support from 
CONACyT (Mexico),
the A. P. Sloan Foundation, 
the Research Corporation,
and the Alexander von Humboldt Foundation.

\begin{appendix}

\section{APPENDIX: ANGULAR MOMENTS}

\label{sc::appendix}

\begin{figure}[!htb]
\begin{center}
\includegraphics[height=7.3cm]{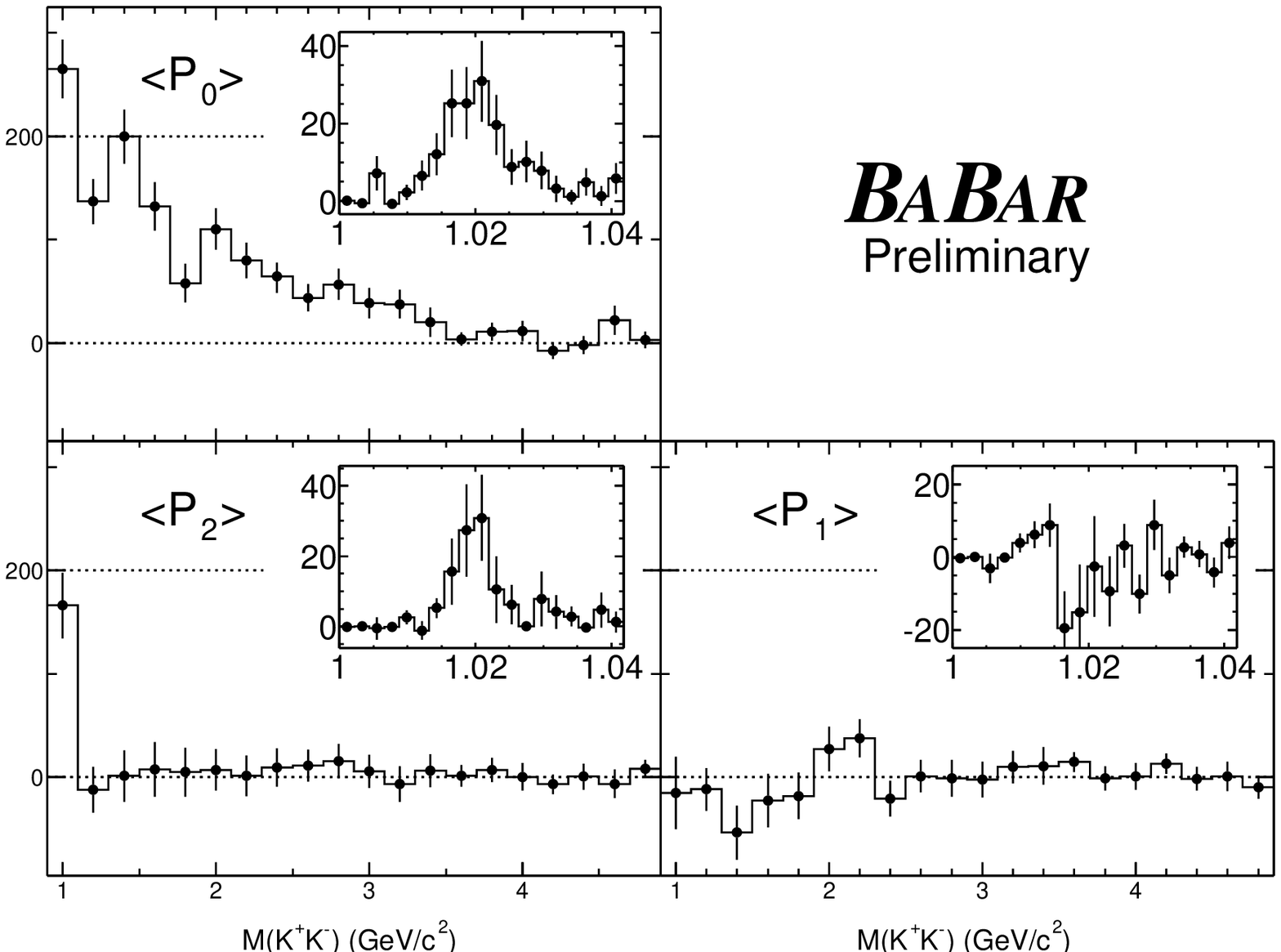}
\includegraphics[height=7.3cm]{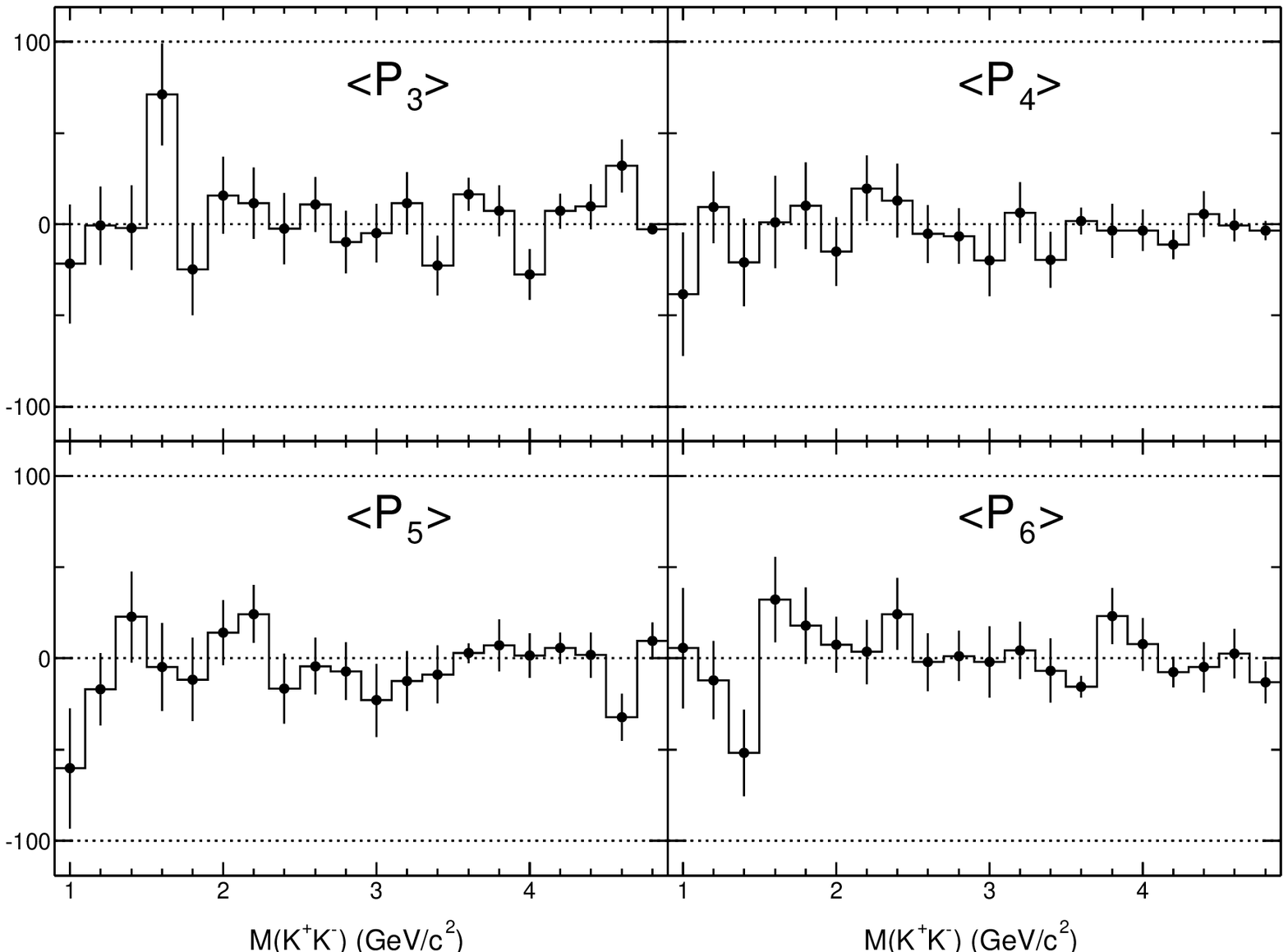}
\caption{Distributions of angular moments $\left < P_l \right >$ with $l=0, \dots, 6$. 
Insets show moments in the $\phi$ mass region.}
\label{fg::moments}
\end{center}
\end{figure}

In the analysis we use the following definitions for the Legendre polynomials
\begin{eqnarray}
	P_0 	&=&	\frac{1}{\sqrt{2}}, \nonumber \\
	P_1	&=&	\sqrt{\frac{3}{2}}   \cos\theta_H,  \nonumber \\
	P_2	&=&	\sqrt{\frac{5}{8}}   \left ( 3 \cos^2\theta_H - 1 \right ), \nonumber \\
	P_3 	&=&	\sqrt{\frac{7}{8}}   \left (  5 \cos^3\theta_H -   3 \cos\theta_H                            \right ), \nonumber \\
	P_4	&=&	\sqrt{\frac{9}{128}} \left ( 35 \cos^4\theta_H -  30 \cos^2\theta_H +   3                    \right ), \nonumber
%	P_5	&=&	\sqrt{\frac{11}{2}} \frac{1}{8}  \left ( 63 \cos^5\theta_H -  70 \cos^3\theta_H +  15 \cos\theta_H       \right ) \nonumber \\
%	P_6	&=&	\sqrt{\frac{13}{2}} \frac{1}{16} \left (231 \cos^6\theta_H - 315 \cos^4\theta_H + 105 \cos^2\theta_H - 5 \right ) \nonumber 	
\end{eqnarray}
which are orthogonal and normalized to unity
$
	\int d(\cos\theta_H) P_l P_{l'} = \delta_{ll'}
$.  
Moments of Legendre polynomials 
$
\left <P_l \right > = 	\int d(\cos\theta_H) P_l(\cos\theta_H) |{\cal A}|^2
$
can be extracted by replacing the integration over the unknown amplitude $\cal A$ with a sum over
signal weights as shown in Eq.~(\ref{eq::mom_sPlot}).
The moments can be related with wave amplitudes $A_{s,p,d}$ and interference phases $\phi_{p,d}$ as follows
\begin{eqnarray}
	\left <P_0 \right >	&=&	\frac{A_s^2 + A_p^2 + A_d^2}{\sqrt{2}}, \nonumber  \\
	\left <P_1 \right >	&=&	\sqrt{2} A_s A_p \cos{\phi_p} ~+~ \sqrt{\frac{8}{5}} A_p A_d \cos(\phi_p-\phi_d),  \nonumber  \\
	\left <P_2 \right >	&=&	\sqrt{\frac{2}{5}} A_p^2 ~+~ \frac{\sqrt{10}}{7} A_d^2  ~+~  \sqrt{2} A_s A_d \cos\phi_d, \nonumber \\
	\left <P_3 \right >	&=&	\sqrt{\frac{54}{35}} A_p A_d \cos(\phi_p - \phi_d),\nonumber  \\
	\left <P_4 \right >	&=&	\frac{\sqrt{18}}{7} A_d^2, \nonumber 
\end{eqnarray}
where we kept only terms relevant for the S, P and D waves (in the nominal result we set $A_d=0$).
Lowest moments plotted in Figure~\ref{fg::moments} are used in the extraction of the P-wave fraction. 
Some of the higher moments  shown in Figure~\ref{fg::moments} are  used for systematic studies.
In the normalization we assume that $\sqrt{2}\left <P_0 \right >$ equals the total number of signal events.

\end{appendix}


\begin{thebibliography}{99}
\bibitem{charge}
Charge-conjugate states are included unless explicitly stated otherwise.
\end{thebibliography}
\end{document}